\documentclass[onecolumn,superscriptaddress,amssymb,amsmath,nobibnotes,aps,prd,showpacs,nofootinbib]{revtex4}%
\pdfoutput=1

\usepackage{graphicx,float}
\usepackage{epsf}
\usepackage{bm}
\usepackage{amsmath,hyperref}
\usepackage{amsfonts}
\usepackage{amssymb}
\usepackage{epstopdf}
\usepackage{natbib}%
\usepackage{rotating}
\usepackage{pdflscape}
\usepackage{adjustbox}

\usepackage[toc,page]{appendix}
\providecommand{\U}[1]{\protect\rule{.1in}{.1in}}
\newcommand{\be}{\begin{equation}}
\newcommand{\ee}{\end{equation}}

\newcommand{\mincir}{\raise
-3.truept\hbox{\rlap{\hbox{$\sim$}}\raise4.truept\hbox{$<$}\ }}
\newcommand{\magcir}{\raise
-3.truept\hbox{\rlap{\hbox{$\sim$}}\raise4.truept\hbox{$>$}\ }}

\usepackage{color}
\definecolor{darkgreen}{rgb}{0., 0.65, 0.1}
\begin{document}

\title{Cosmological bounds on dark matter-photon coupling}

\author{Suresh Kumar}
\email{suresh.kumar@pilani.bits-pilani.ac.in}
\affiliation{Department of Mathematics, BITS Pilani, Pilani Campus, Rajasthan-333031,India}
 
\author{Rafael C. Nunes}
\email{rafadcnunes@gmail.com}
\affiliation{Departamento de F\'isica, Universidade Federal de Juiz de Fora, 36036-330,
Juiz de Fora, MG, Brazil}

\author{Santosh Kumar Yadav}
\email{sky91bbaulko@gmail.com}
\affiliation{Department of Mathematics, BITS Pilani, Pilani Campus, Rajasthan-333031,India}

\pacs{95.35.+d; 95.36.+x; 14.60.Pq; 98.80.Es}

\begin{abstract}
We investigate an extension of the $\Lambda$CDM model where the dark matter (DM) is coupled to photons, inducing a 
nonconservation of the numbers of particles for both species, where the DM particles are allowed to dilute throughout
the cosmic history with a small deviation from the standard evolution decaying into photons, while the associated
scattering processes are assumed to be negligible.
In addition, we consider the presence of massive neutrinos with the effective number of species $N_{\rm eff}$ as a free parameter. 
The effects of the DM-photon coupling on the cosmic microwave background (CMB) and matter power spectra are analyzed. 
We derive the observational constraints on the model parameters
by using the data from CMB, baryonic acoustic oscillation (BAO) measurements, the recently measured new 
local value of the Hubble constant from the Hubble Space Telescope, and large scale structure (LSS) 
information from the abundance of galaxy clusters. The DM-photon coupling parameter $\Gamma_{\gamma }$  
is constrained to $\Gamma_{\gamma } \leq 1.3 \times10^{-5}$ (at 95\% C.L.)  from the joint analysis carried out by using all the mentioned data sets.
The neutrino mass scale $\sum m_{\nu}$ upper bounds at 95\% C.L. are obtained as $\sum m_{\nu} \sim 0.9$ eV and $\sum m_{\nu} \sim 0.4$ eV with 
and without the LSS data, respectively. We observe that the DM-photon coupling cause significant changes in the best fit value of $N_{\rm eff}$ 
but yields statistical ranges of $N_{\rm eff}$ compatible with the standard predictions, and we do not find any evidence of dark radiation. 
Due to nonconservation of photons in our model, we also evaluate and analyze the effects on the BAO acoustic scale at the drag epoch.
The DM-photon coupling model yields high values of Hubble constant consistent with 
the local measurement, and thus alleviates the tension on this parameter.
\end{abstract}

\maketitle

\section{Introduction}
\label{sec:intro}

The nature of dark matter (DM) is one of the most important open questions in modern science. 
There is much indirect evidence for the existence of DM (see \cite{DM01,DM02} for review), but little is known about its properties 
(spin, parity, mass, interaction cross section, lifetime etc). From the point of view of cosmological and astrophysical observations, 
it does not require the DM to be absolutely stable but only with a lifetime longer than the age of the Universe, in general. 
The search for decaying DM can be used to impose upper bounds on the decay width of DM into different final states; 
see \cite{DM03} for review of decaying DM signals in cosmic ray antimatter, gamma rays and neutrinos. 
Searches for DM decay have been intensively investigated from IceCube telescope data \cite{IceCube}. 
The DM decay into photons (and into photon and neutrino) has been considered and constrained from cosmic ray emission in \cite{Dgamma01,Dgamma02,Dgamma03}. 
Also, the possibility of late-time DM decay has been studied in the context of the problems related to the structure formation in the Universe at small scales \cite{DM05,DM06}.



On the other hand, a DM-photon interaction via an elastic scattering has been considered in \cite{DM07, DM07_2} (see also \cite{DM08} for recent developments), 
and an upper bound $\sigma_{\rm DM-\gamma} \lesssim 10^{-32} \, (m_{\rm DM}/{\rm GeV}) \, \rm cm^2$ for the elastic scattering cross section is derived 
in \cite{DM09}. An elastic scattering cross section $\sigma_{\rm DM-\nu} \lesssim 10^{-33} \, (m_{\rm DM}/{\rm GeV})\, {\rm cm^2} $ for DM-neutrino, and
$\sigma_{\rm DM-DE} \lesssim 10^{-29} \, (m_{\rm DM}/{\rm GeV})\, {\rm cm^2} $ for the DM-dark energy, are presented in \cite{DM10} and \cite{DM11}, 
respectively. Recently, many studies \cite{DDM01,DDM02,DDM03,DDM04,DDM05,DDM05_1,DDM05_2} have been carried out where the DM
interacts with a thermal background of dark radiation \footnote{Dark radiation is inferred by stating that the radiation content of the
Universe is not only due to photons and neutrinos but also due to some 
extra relativistic relics (see \cite{eu1} for recent constraints on the total effective number of relativistic species and \cite{Neff01} 
for a recent review).}. Possible evidence for DM-dark radiation
interaction is reported in \cite{DDM06}. Also, 
DM decay models have been investigated in order to solve or assuage the problems associated with the standard model 
\cite{DDM01,DDM02,DDM03,DDM04,DDM05,DDM06}.


Despite the success of the $\Lambda$CDM model, where DM particles interact only gravitationally with other particles, 
some free parameters of this model are currently in tension with some observational estimates. The most well-known tensions are in the 
estimation of the Hubble constant $H_0$ and the amplitude of matter density fluctuations $\sigma_8$ from Planck cosmic microwave background (CMB) data and direct measurements 
(see Sec. \ref{results} for more details). However, these tensions may be the outcome of systematic effects in data rather than a hint of
new physics beyond $\Lambda$CDM. For instance, the authors in \cite{H_01} argue that the supernova measurements of $H_0$ are overestimated 
 due to the local environment bias in supernovae type Ia standardized magnitudes.
On the other hand, the findings in \cite{H_02} suggest that the tension in $H_0$ distance ladders is likely not a result of supernova systematics
that could be expected to vary between optical and near-infrared wavelengths, like dust extinction.
Likewise, the tension on $\sigma_8$ has at least two sources: i) galaxy cluster counts, ii) weak lensing. 
Regarding the galaxy cluster counts, the ways to alleviate the $\sigma_8$ tension have been discussed in \cite{sigma2}. 
Furthermore, the new results on the weak lensing by the Dark Energy Survey (DES) collaboration \cite{sigma3} exhibit no tension in
the $\sigma_8$-parameter measurement. The possible systematic effects in the discrepancy between
CMB and large scale structures (LSS) are also discussed in \cite{sigma4}.




In the present paper, we consider a cosmological model with a nonminimal DM-photon coupling in which the interaction
is assumed to lead to a scenario where the DM decays into photons. This phenomenological scenario of DM-photon coupling can be justified for a possible ``dark electromagnetism,'' 
as proposed initially in \cite{DMDR} for DM-dark radiation coupling.
Our main aim in this work is to investigate the observational constraints on this cosmological scenario using the  Planck CMB data, baryonic acoustic oscillation (BAO) measurements, 
the recently measured new local value of the Hubble constant from the Hubble Space Telecsope (HST) and LSS data from the abundance of galaxy clusters. 
We find that this possible nonconservation of the energy density of photons, even though it is too small (coupling parameter $\sim 10^{-5}$), 
can be a natural candidate to solve the $H_0$ tension as discussed in Sec. \ref{results}.


The paper is organized as follows: In the next section, we introduce the DM-photon coupling model, and 
we discuss the quantitative effects on CMB and matter power spectra due to the DM-photon coupling. In Sec. \ref{results}, we derive observational boundaries on a possible
DM-photon interaction. The final section includes the conclusion of the study. 
In what follows, a subindex
0 attached to any parameter means the value of the parameter at the present time. An over dot and prime represent the cosmic time
 and conformal time derivatives, respectively.

\section{Dark matter-photon coupling model}

The background evolution energy density $\rho_{\rm ddm}$ of the decaying cold DM follows the standard line already well known in the literature, 
where a nonconservation of the number density of DM particles leads to nonconservation of the energy-momentum tensor of the DM particles, 
$\nabla_{\mu} T^{\nu \mu}_{\rm ddm} = Q$. Here the coupling function $Q$ accounts for the decay of DM. 
The index ${\rm ddm}$ represents decaying DM. In the present study, we consider that DM can decay into photons. 
Thus, the background density equations, assuming Friedmann-Robertson-Walker (FRW) Universe, take the form

\begin{align}
\label{ddm_fundo}
 \rho'_{\rm ddm} + 3 \frac{a'}{a} \rho_{\rm ddm} = - \frac{a'}{a} \Gamma_{\gamma} \rho_{\rm ddm}, 
\end{align}
\begin{align}
\label{gamma_fundo}
 \rho'_{\gamma} + 4 \frac{a'}{a} \rho_{\gamma} = \frac{a'}{a} \Gamma_{\gamma} \rho_{\rm ddm}, 
\end{align}
where $\Gamma_{\gamma}$  
is a dimensionless parameter characterizing the coupling between DM and photons, and prime denotes the conformal time derivative. The quantity $\rho_{\gamma}$ is the energy density of photons. 
It is known that the energy-momentum conservation equation of $i$th coupled fluid in a cosmological scenario 
reads as $\nabla_{\mu} T^{\nu \mu}_{\rm i} = Q^{\nu}_i$ with 
$\sum_i Q^{\nu}_i = 0$. We notice that Eqs. (\ref{ddm_fundo}) and (\ref{gamma_fundo}) satisfy this condition
with $Q_{\rm ddm} = - a'/a \, \Gamma_{\gamma} \rho_{\rm ddm}$ and $Q_{\gamma} = a'/a \, \Gamma_{\gamma} \rho_{\rm ddm}$, respectively. 
We adopt $\Gamma_{\gamma} > 0$ in order to have a decaying DM along the cosmic expansion. Usually the DM decay rate is considered constant. But in principle, it could be time variable as well. So without loss of generality, the decay rate can be defined as $\Gamma = \Gamma_\gamma \mathcal{H}/a$, where $\mathcal{H}=\frac{a'}{a}$ is conformal Hubble parameter.


Solving (\ref{ddm_fundo}) and (\ref{gamma_fundo}), we find

\begin{align}
 \rho_{\rm ddm} = \rho_{\rm ddm0}a^{-3 -\Gamma_{\gamma}},
\end{align}


\begin{align}
 \rho_{\gamma}=\rho_{\gamma0} a^{-4} + \frac{\Gamma_{\gamma}}{1- \Gamma_{\gamma}} \rho_{\rm ddm0}(a^{-3 -\Gamma_{\gamma}}-a^{-4}),
\end{align}
where for $\Gamma_{\gamma} = 0$, we recover the standard evolution equations for the DM and photons. 

\subsection{Perturbation equations}

Now we consider the evolution of linear cosmological perturbations in our model. In the synchronous gauge, the 
line element of the linearly perturbed FRW metric is given by

\begin{align}
 ds^2 = - a^2d\tau^2 + a^2[(1- 2 \eta)\delta_{ij} + 2 \partial_i \partial_j E]dx^idx^j,
\end{align}
where $k^2 E = -2/h -3 \eta$, restricting to the scalar modes $h$ and $\eta$.

Using $\nabla_{\mu} T^{\nu \mu}_{\rm i} = Q^{\nu}_i$, the continuity and Euler equations of the $i$th coupled fluid, given the above metric, 
are written as 

\begin{eqnarray}
 \delta'_i + 3\mathcal{H} (c^2_{s,i} -w_i)\delta_i + 9\mathcal{H}^2 (1+w_i) (c^2_{s,i} - c^2_{a,i}) \frac{\theta_i}{k^2}  
 + (1+w_i)\theta_i -3(1+w_i)\eta' \nonumber \\
 + (1+w_i) \left(\frac{h'}{2} + 3\eta' \right)  = \frac{a}{\rho_i}(\delta Q_i - 
 Q_i \delta_i) + a \frac{Q_i}{\rho_i} \Big[3\mathcal{H} (c^2_{s,i} - c^2_{a,i}) \Big] \frac{\theta_i}{k^2},
\end{eqnarray}

\begin{eqnarray}
\theta'_i + \mathcal{H} (1 - 3c^2_{s,i})\theta_i - \frac{c^2_{s,i}}{(1+w_i)} k^2 \delta_i 
= \frac{a Q_i}{(1 + w_i)\rho_i} \Big[\theta_{\rm ddm} - (1+c^2_{s,i})\theta_i \Big],
\end{eqnarray}
where we have chosen the momentum transfer in the rest frame of DM. Here, $w_i$, $c^2_{a,i}$, and $c^2_{s,i}$ are the 
equation of state, adiabatic sound speed, and physical sound speed in the rest frame of the $i$th fluid, respectively.
As expected, for $Q_i=0$ in the above equations, we obtain the standard continuity and Euler equations of the $i$th fluid. 
This methodology was initially used to describe the linear perturbations of a dark sector interaction between DM and dark energy (see \cite{lixin14} 
and references therein).

The next step is to particularize the fluid approximation equations to the coupled system of DM and photon. We have,

\begin{eqnarray}
\label{delta_gamma}
\delta'_{\gamma} + \frac{4}{3}\theta_{\gamma} + \frac{2}{3}h' = a \Gamma_{\gamma} \mathcal{H} \frac{\rho_{\rm ddm}}{\rho_{\gamma}} (\delta_{\rm ddm} - \delta_{\gamma}),
\end{eqnarray}

\begin{eqnarray}
\label{theta_gamma}
 \theta'_{\gamma} - \frac{1}{4}k^2 (\delta_{\gamma} - 4\sigma_{\gamma}) - a n_e \sigma_T (\theta_b - \theta_{\gamma}) 
 = \frac{3}{4} a \Gamma_{\gamma} \mathcal{H} \frac{\rho_{\rm ddm}}{\rho_{\gamma}} (\theta_{\rm ddm} - \frac{4}{3} \theta_{\gamma}),
\end{eqnarray}
describing the continuity and Euler equations for photons, respectively.

Lastly, the DM evolution is given by

\begin{eqnarray}
 \delta'_{\rm ddm} + \frac{h'}{2} = 0.
\end{eqnarray}

In (\ref{theta_gamma}), $\theta_b$ is the divergence of baryon fluid velocity, where the term $a n_e \sigma_T (\theta_b - \theta_{\gamma})$ 
is due to the collision term
before recombination between photons and baryons, which are tightly coupled, interacting mainly via Thomson scattering. 
The Euler equation derived for DM in the synchronous gauge reads $\theta_{\rm ddm} = 0$. 


Recently, the authors in \cite{DM08} have considered an elastic scattering between DM and photons, and described the complete treatment
to Boltzmann hierarchy for photons, including changes in expansion for $l \geq 3$. 
Here, we are considering that the interaction between DM and photons is interpreted for a nonconservation in the numbers of particles for both species, 
where the DM particles can undergo dilution throughout the cosmic history with a small deviation from the standard evolution 
decaying into photons. Thus, the process here is different from the DM-photon elastic scattering interaction, where
the  particle number  density is always conserved, and changes do not occur at background level as well as to continuity equations of 
the DM and photons. Beyond the changes in the Euler and continuity equations (also in the background dynamics), we believe that a complete treatment 
of Boltzmann hierarchy for $l \geq 3$ must also be carried out in our model, 
but this is beyond our goal in the first steps of the present investigation. We hope to develop it in another future communication.

\subsection{Effects of the dark matter-photon coupling on the CMB TT and matter power spectra}\label{powerspectra}

It is well known that a nonconservation in the photon number density can affect the anisotropy of the CMB. 
Photon production (or destruction) has been considered 
in other contexts \cite{CMB_gamma01, CMB_gamma02, CMB_gamma03, CMB_gamma04, CMB_gamma05, CMB_gamma06, CMB_gamma07, CMB_gamma08}.
In general, a change in the standard dynamics of photons can affect CMB and another important cosmological relationships in various ways,
like the CMB spectral distortions, secondary CMB anisotropies, luminosity distance etc. 
Here, we focus on the background and perturbative changes (as described in this section) in order to test a cosmological scenario with 
decay of DM  into photons.


%

\begin{figure*}[h] 
\includegraphics[width=7.5cm]{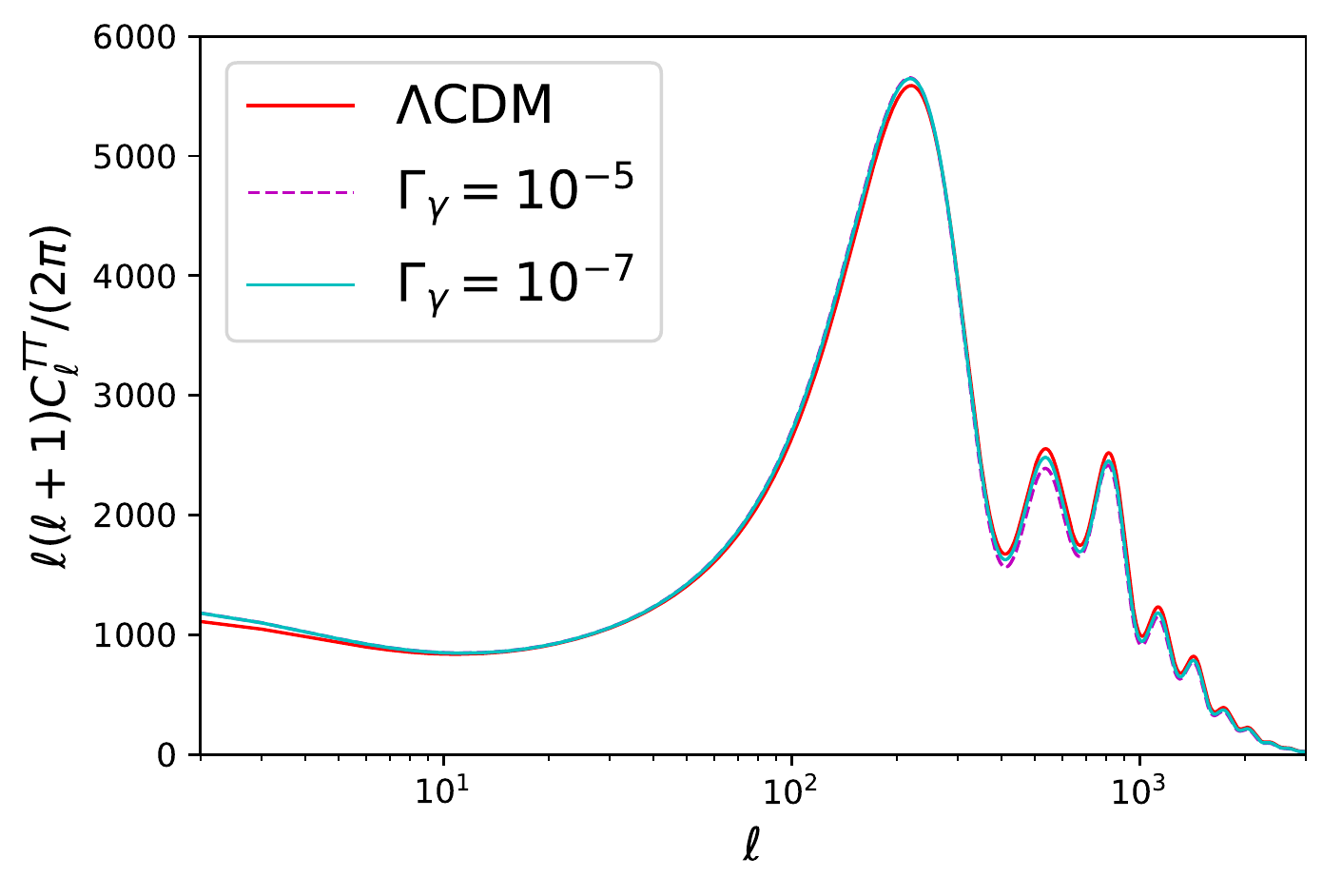}
\includegraphics[width=7.5cm]{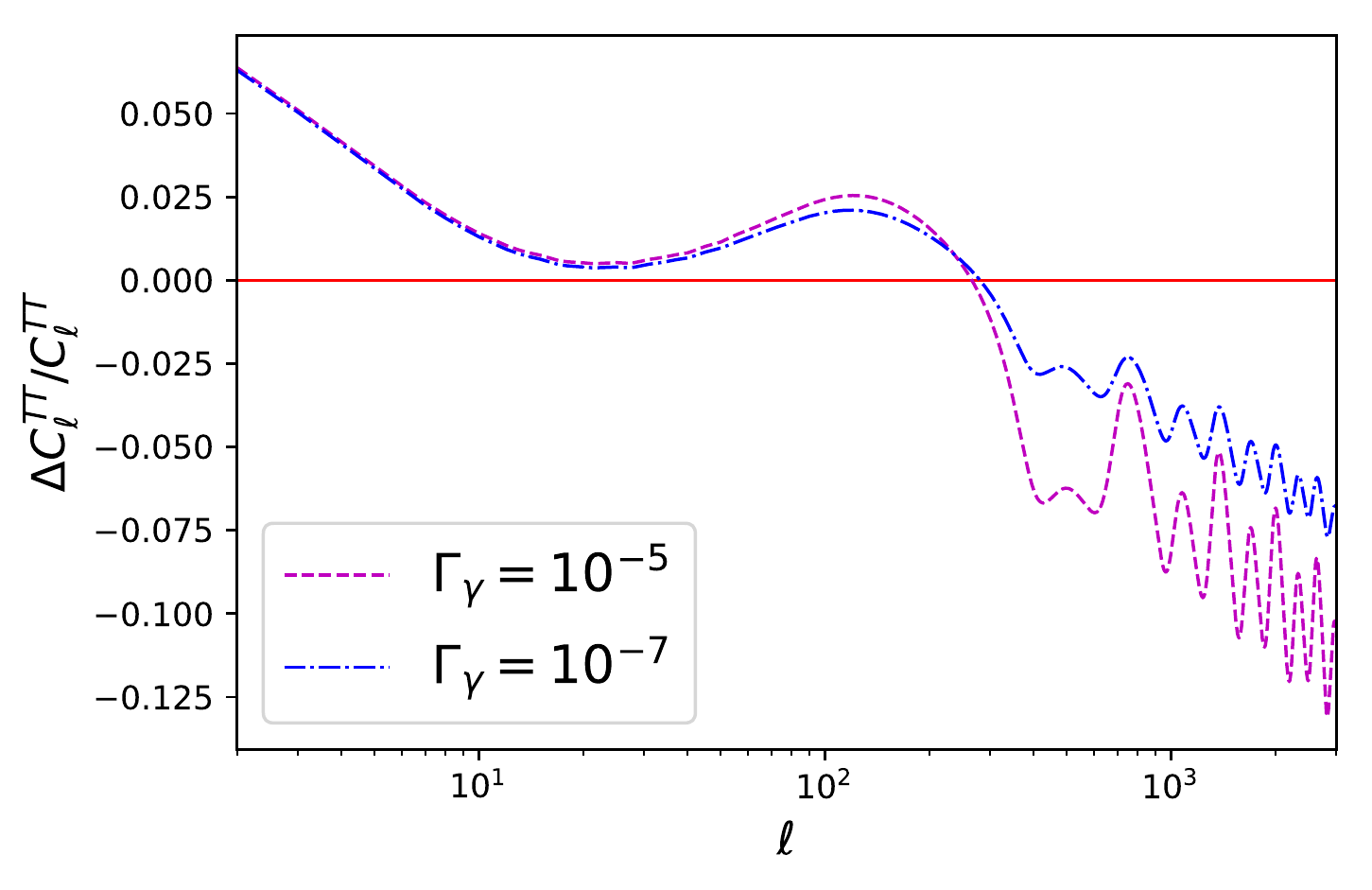}
\caption{Theoretical prediction and relative deviations of the CMB TT power spectrum from the base
line Planck 2015 $\Lambda$CDM model for some values of $\Gamma_{\gamma}$ while
the other parameters are fixed to their best-fit mean values as given in
Table \ref{Table_M1}.} \label{fig1a}
 \end{figure*}
\begin{figure*}[h] 
 \includegraphics[width=7.5cm]{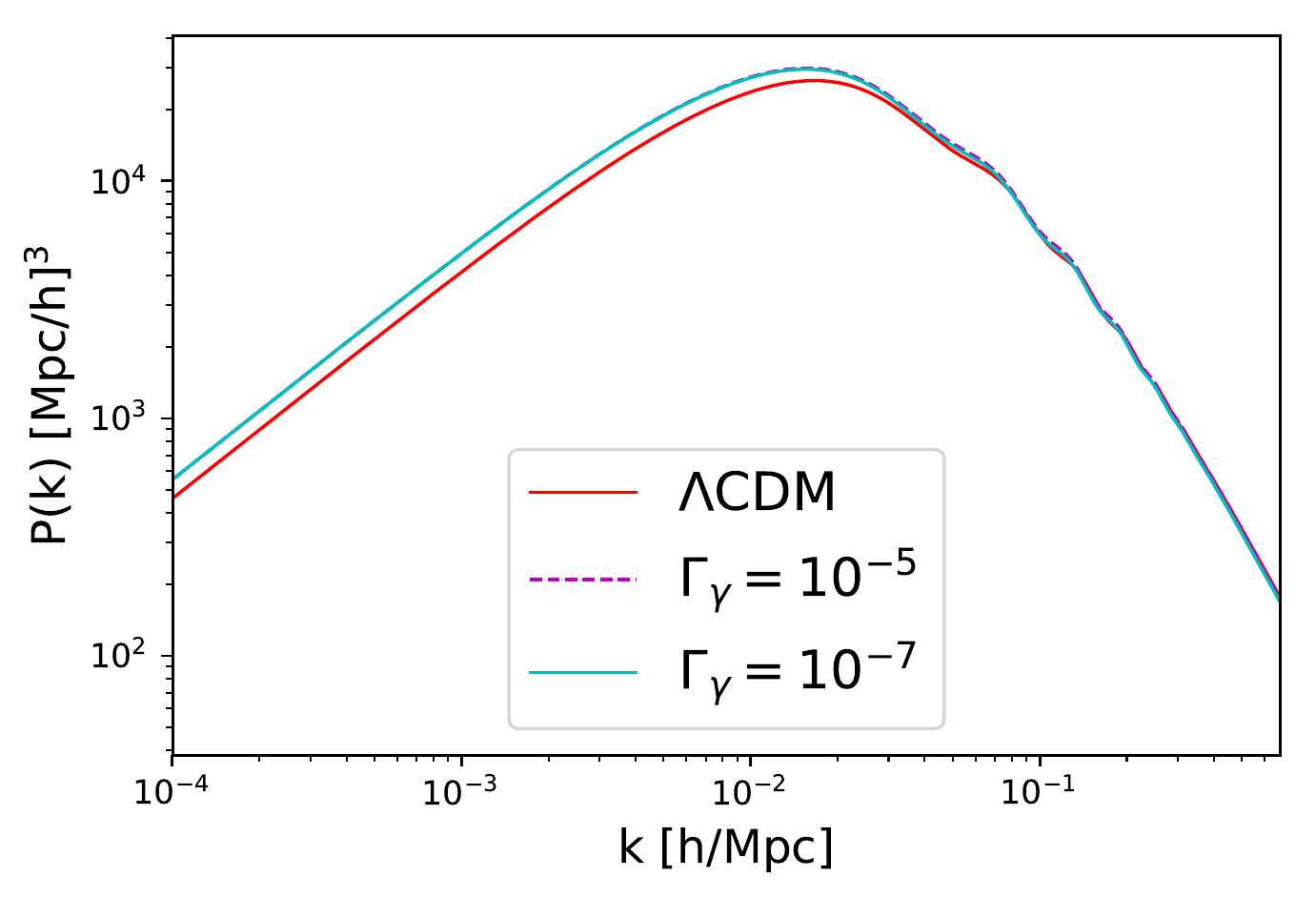}
\includegraphics[width=7.5cm]{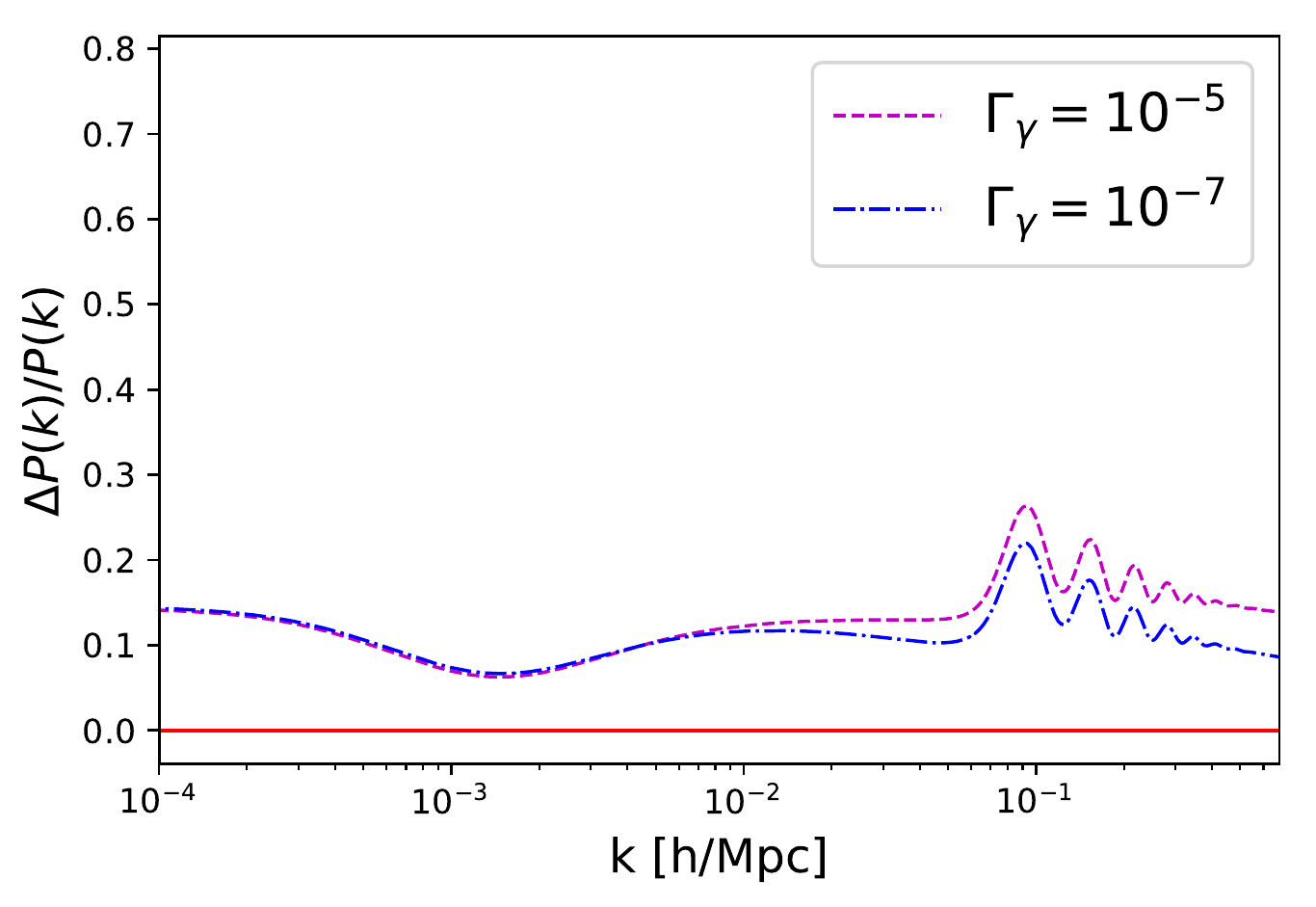}
\caption{Theoretical prediction and relative deviations of the matter power spectrum from the base
line Planck 2015 $\Lambda$CDM model for some values of $\Gamma_{\gamma}$ while
the other parameters are fixed to their best-fit mean values as given in
Table \ref{Table_M1}.}\label{fig3a}
 \end{figure*}

Figures \ref{fig1a} and \ref{fig3a} show the theoretical predictions of the CMB temperature (TT) and matter power specta at $z=0$, as well as 
the relative deviations from the base line Planck 2015 $\Lambda$CDM model, for the coupling parameter in the range 
$\Gamma_{\gamma} \in [10^{-5}, 10^{-7}]$. We see that on large angular scales ($l < 30$) we have deviations approximately until 
5\%  and up to 12\% on small scales on CMB TT. These changes are quantified by the reduction in magnitude of
the acoustic peaks at small scales by collisional damping and the enhancement of the first acoustic peaks due 
to a decrease in the photon diffusion length. The effects on matter power spectrum are about 10\%-15\% on large scale and oscillations 
around 15\%-25\% on small scale (where nonlinear effects may be predominant). These changes on CMB TT and matter power spectra
are similar in order of magnitude or even smaller than in other decaying DM models described in the literature. 
In the next section, we derive the observational constraints on the model under consideration.\\

\section{Observational constraints}
\label{results}

\subsection{Data sets, methodology and the model parameters}

In order to constrain the model under consideration, we use the following data sets. \\

\noindent\textbf{CMB}: A conservative data set comprising of cosmic microwave background temperature power spectrum (TT), low-$l$ polarization and lensing reconstruction 
from Planck 2015 \cite{Planck2015}. \\

\noindent\textbf{BAO}: The baryon acoustic oscillation measurements from the  Six  Degree  Field  Galaxy  Survey  (6dF) \cite{bao1}, 
the  Main  Galaxy  Sample  of  Data  Release 7  of  Sloan  Digital  Sky  Survey  (SDSS-MGS) \cite{bao2}, 
the  LOWZ  and  CMASS  galaxy  samples  of  the Baryon  Oscillation  Spectroscopic  Survey  (BOSS-LOWZ  and  BOSS-CMASS,  
respectively) \cite{bao3},  and the distribution of the Lyman Forest in BOSS (BOSS-Ly) \cite{bao4}, as 
summarized in Table I of \cite{baotot}.\\

\noindent\textbf{HST}: The recently measured new local
value of Hubble constant, $H_0=73.24 \pm 1.74$  km s${}^{-1}$ Mpc${}^{-1}$ by the Hubble Space Telescope, as reported in \cite{riess}. \\

\noindent\textbf{LSS}: Three probes of large scale structure: the constraints, $\sigma_8\left(\frac{\Omega_m}{0.27}\right)^{0.46} = 0.774 \pm 0.040$ 
from the weak lensing survey CFHTLens \cite{LSS1}, $\sigma_8\left(\frac{\Omega_m}{0.27}\right)^{0.30} = 0.782 \pm 0.010$ from Planck Sunyaev-Zeldovich cluster 
mass function \cite{LSS2} and $\sigma_8\left(\frac{\Omega_m}{0.30}\right)^{0.50} = 0.651 \pm 0.058$ from the Kilo Degree Survey KIDS-450 \cite{LSS3}.\\

We have implemented the DM-photon coupling model in the publicly available CLASS \cite{class} code, and used the Metropolis Hastings algorithm in Monte Python \cite{monte} code with uniform priors on the model parameters to obtain correlated Markov Chain Monte Carlo samples by 
considering four combinations of data sets: CMB + BAO, CMB + BAO + HST, CMB + BAO + LSS, and CMB + BAO + HST + LSS.   
We have ensured the convergence of the Monte Carlo Markov Chains for all the model parameters according to the Gelman-Rubin criteria \cite{Gelman_Rubin}.
We have analyzed the output samples by using the GetDist Python package \cite{antonygetdist}. We have chosen the 
minimal data set CMB+BAO because adding BAO data to CMB reduces the error bars on the parameters. The other data set combinations are considered 
in order to investigate how the constraints on various free parameters and derived parameters of the DM-coupling model are affected by the inclusion of HST and LSS data.\\

In general, it is common to consider only the $\Lambda$CDM model as a cosmological scenario to investigate decaying DM models. In our analysis, we consider 
that dark energy is in the form of a perfect fluid with a constant equation of state parameter $w_{\rm de}$, and investigate its possible effects and 
deviations from the $\Lambda$CDM case in the context under study. Also, to the knowledge of the authors, the constraints on $w_{\rm de}$ from the 
cosmological data have not yet been studied in the context of the decaying DM. We have also considered the presence of neutrinos and set the order of mass 
on the normal hierarchy with a minimum sum of neutrino masses to be 0.06 eV. The total effective number of relativistic species ($N_{\rm eff}$) 
is also taken as a free parameter. Finally, the base parameters set for the DM-photon coupling  model is
 \[
 P = \{100\omega_{\rm b}, \, \omega_{\rm cdm}, \, 100\theta_{s}, \, \ln10^{10}A_{s}, \,  
 n_s, \, \tau_{\rm reio}, \, w_{\rm de}, \, \sum m_{\nu}, \, N_{\rm{eff}}, \, \Gamma_{\gamma} \},
 \]
where the first six parameters are also the base parameters of the minimal $\Lambda$CDM model \cite{p13}. 

\subsection{Results and discussion}

Table \ref{Table_M1} summarizes the observational constraints on the free parameters and some derived parameters of the DM-photon coupling model 
for the four combinations of data sets:
CMB + BAO, CMB + BAO + LSS, CMB + BAO + HST and  CMB + BAO + LSS + HST.

\begin{table*}[!ht] 
\caption{\label{Table_M1} {Constraints on the free parameters and some derived parameters of the DM-photon coupling model for four combinations of data sets. 
The upper and lower values with mean value of each parameter denote 68\% C.L. and 95\% C.L. errors. 
For $\sum m_{\nu}$ and $\Gamma_\gamma$, the upper bounds at 95\% C.L. are mentioned. The parameter $H_0$ is measured in the units of 
km s${}^{-1}$ Mpc${}^{-1}$, $r_{\rm{drag}}$ in Mpc, whereas $\sum m_{\nu} $ is in the units of eV.}}
\resizebox{\textwidth}{!}{%
\begin{tabular} { |l|l|l|l|l|}  \hline  & & & & \\[-1.7ex]
 Parameter &  CMB + BAO   & CMB + BAO + LSS & CMB + BAO + HST & CMB + BAO + LSS + HST\\  [0.7ex]
\hline
& & & & \\[-1.8ex] $10^{2}\omega_{\rm b }$ &$2.29^{+ 0.14+0.27}_{-0.14-0.26}$ & $2.34^{+0.19+0.31}_{-0.16-0.34}  $ &
$2.33^{+ 0.09+0.19}_{-0.09-0.18}  $     & $2.31^{+0.09+0.20}_{-0.09 -0.18}   $\\[1ex]

$\omega_{\rm cdm }  $ &  $0.127^{+0.013+0.024}_{-0.013-0.025}            $ & $0.129^{+0.018+0.028}_{-0.015-0.030}   $ &$0.130^{+0.007+0.016}_{-0.008-0.015}$& $0.125^{+0.007+0.017}_{-0.008-0.015}$ \\[1ex]

$\ln10^{10}A_{s }$& $3.107^{+0.034+0.082}_{-0.045-0.075}   $ &$3.127^{+0.045+0.087}_{-0.045-0.084}            $     & $3.108^{+0.038+0.080}_{-0.045-0.077}   $ &$3.130^{+ 0.043+0.085}_{-0.043-0.084}            $\\[1ex]

$100 \theta_{s }  $ & $1.0412^{+0.0009+0.0021}_{-0.0011-0.0019}$&$1.0409^{+0.0009+0.0023}_{-0.0014-0.0019}$
&   $1.0409^{+0.0007+0.0015}_{-0.0007-0.0014}        $&  $1.0411^{ +0.0007+0.0016}_{-0.0007-0.0015}        $\\[1ex]

$n_{s }         $ & $0.977^{+0.012+0.028}_{-0.013-0.024}   $&$0.978^{+0.013+0.025}_{-0.013-0.024}            $  & $0.977^{+0.013+0.026}_{-0.013-0.024}            $  &  $0.980^{+0.012+0.025}_{-0.012-0.024}            $\\[1ex]

$\tau_{\rm reio }   $ &  $0.084^{+0.017+0.038}_{-0.021-0.036}   $ &$0.096^{+0.019+0.041}_{-0.022-0.038}   $  & $0.084^{+0.018+0.038}_{-0.021-0.037}    $& $0.098^{+ 0.020+0.040}_{-0.020-0.039}            $    \\[1ex]

$w_{\rm{de}}    $ & $-1.03^{+0.10+0.18}_{-0.08-0.19}    $&$-1.15^{+0.12+0.20}_{-0.10-0.22}                $   & $-1.03^{+0.11+0.18}_{-0.08-0.20}    $ &$-1.13^{+0.11+0.19}_{-0.09-0.20}    $     \\[1ex]

$\sum m_{\nu}      $ & $<0.38$ &  $< 0.97$   &  $<0.37   $ &   $ <0.83 $\\ [1ex]

$N_{\rm{eff}}   $ & $3.51^{+0.43+0.82}_{-0.43-0.84}              $  &$3.65^{+0.61+0.90}_{-0.46-1.00}        $  & $3.60^{+0.34+0.69}_{-0.34-0.63}              $ & $3.58^{+ 0.36+0.72}_{-0.36-0.68}              $\\[1ex]


$\Gamma_{\gamma }$ & $<2.3\times10^{-5}$ & $<2.6\times 10^{-5}$ & $<9.0\times10^{-6}   $ &$<1.3 \times10^{-5}   $\\ [1ex]

\hline

& & & & \\[-1.7ex]
$\Omega_{\rm{m} }$ & $0.293^{+0.013+0.026}_{-0.013-0.026}            $  &$0.284^{+ 0.015+0.028}_{-0.015-0.029}            $& $0.293^{+0.013+0.025}_{-0.013-0.024}            $  &$0.287^{+ 0.013+0.025}_{-0.013-0.026}            $   \\[1ex]
  
  $H_{\rm 0}$ & $71.9^{+4.0 +7.8}_{-4.0-7.1}              $
 &$74.7^{+ 5.1+10.0}_{-5.1-10.0}               $ & $72.8^{+ 1.7+3.2}_{-1.7-3.3}               $  &   $73.3^{+1.7+3.5}_{-1.7-3.3}               $\\[1ex]
 
  $\sigma_8       $ & $0.829^{+0.020+0.040}_{-0.020-0.038}            $&$0.781^{+ 0.015+0.028}_{-0.015-0.028}            $ & $ 0.829^{+ 0.019+0.037}_{-0.019-0.037}    $         &$0.778^{+0.014+0.027}_{-0.014-0.027}            $\\[1ex]
  
 $r_{\rm{drag} }$ & $142.5^{+6.8+15}_{-8.2-13}       $&$140.5^{+6.1+20.0}_{-11.0-10.0}        $  & $140.5^{+ 4.4+8.6}_{-4.4-8.5}              $
   &  $141.9^{+ 4.7+9.4}_{-4.7-9.1}              $\\[1ex]
  
  \hline
\end{tabular}}
\end{table*}

\vspace{1cm}

\begin{figure}[h] 
 \includegraphics[width=7.5cm]{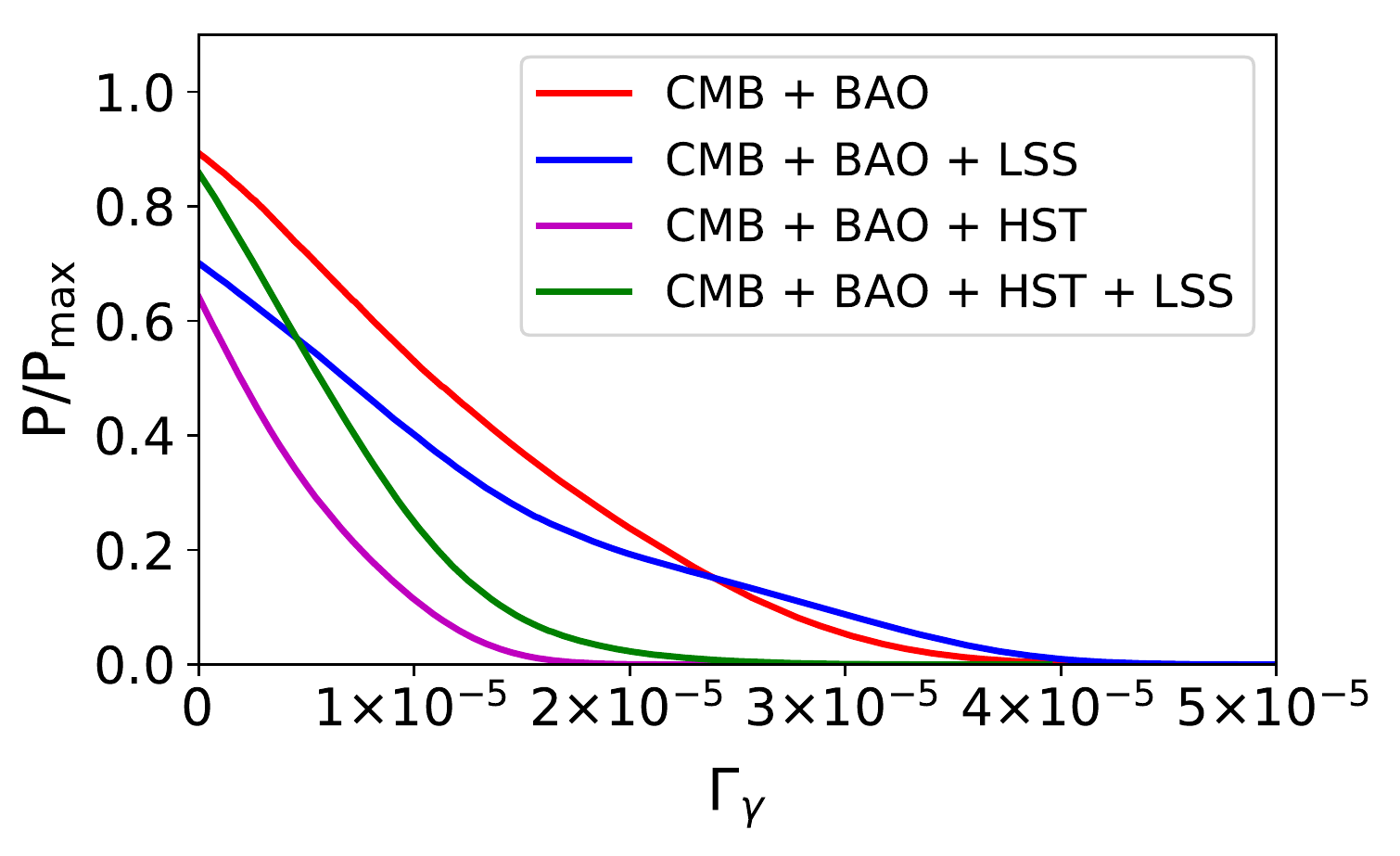}
  \caption{\label{Gamma_model1} {One-dimensional marginalized distribution for $\Gamma_{\gamma}$.}}
\end{figure}

We note that with all the data sets analyzed here, the DM-photon coupling parameter $\Gamma_{\gamma }$ is very small with the order $10^{-5}$. 
 We have $\Gamma_{\gamma } \leq 1.3 \times10^{-5}$ at 95\% C.L. from the joint analysis using full
data CMB + BAO  + LSS + HST. Also, see Fig. \ref{Gamma_model1} which shows the one-dimensional marginalized distribution for $\Gamma_{\gamma}$.
The coupling parameter has the same order of magnitude in all four cases, but a smaller amplitude is observed in the case CMB + BAO + HST.

Even though it is very small (reasonably expected), the nonconservation of photons due to the DM-photon coupling leads to 
significant effects on the CMB, which can directly affect the other cosmological parameters in particular $H_0$ and $\sigma_8$. 
Further, the constraints on $w_{\rm de}$ with CMB + BAO  + LSS  and full data CMB + BAO  + LSS + HST are $w_{\rm de}=-1.15^{+0.12}_{-0.10}$ and 
$w_{\rm de}=-1.13^{+0.11}_{-0.09}$ both at 68\% C.L., respectively. 
Therefore, a phantom behavior of dark energy is minimally favored in the DM-photon coupling model when LSS data are included in the analysis. 
In the other two cases without LSS data, the dark energy behavior is similar to the cosmological constant since $w_{\rm de}\sim -1$.  
 

\begin{figure*}
\centering
\includegraphics[width=12.0cm]{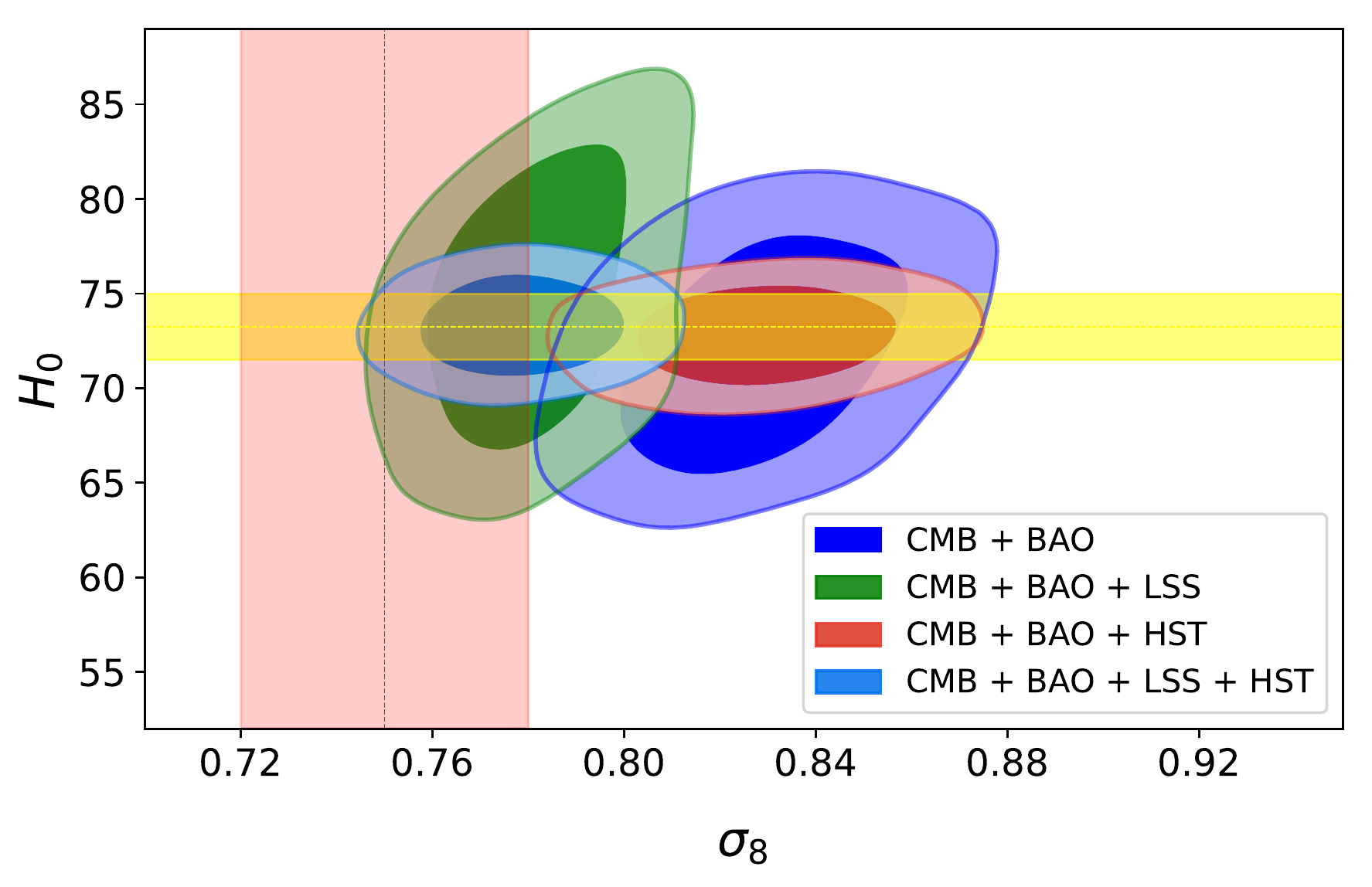}
\caption{\label{PS_model1} {68\% C.L. and 95\% C.L. regions 
for $\sigma_8$ and $H_0$. The horizontal yellow band  corresponds to $H_0=73.24\pm 1.74$ km s${}^{-1}$ Mpc${}^{-1}$ whereas the vertical light red 
band corresponds to $\sigma_8= 0.75\pm 0.03$.  }}
\end{figure*}

Assuming a base line standard $\Lambda$CDM model, the value of Hubble constant $H_0 = 66.27\pm 0.66 $ km s ${}^{-1}$ Mpc${}^{-1}$ from 
Planck CMB measurement \cite{Planck2015} is about $3\sigma$ lower than the current local measurement $H_0 = 73.24\pm 1.74 $ km s${}^{-1}$ Mpc${}^{-1}$ as 
reported in \cite{riess}. This discrepancy between both measures has been widely discussed in the 
literature as a possible indication of physics beyond the minimal $\Lambda$CDM model. 
The interaction between DM and photons gives rise to a very weak ``DM drag'' which damps the growth of matter density perturbations throughout radiation 
domination, and therefore can act to reconcile the tension on $H_0$ between predictions from the CMB and direct measurements. A similar scenario also arises in DM-dark radiation interaction models. Figure \ref{PS_model1} shows the parametric space of $H_0$ - $\sigma_8$, where
the horizontal yellow band corresponds to  $H_0=73.24\pm 1.74$ km s${}^{-1}$ Mpc${}^{-1}$.
From Table \ref{Table_M1}, we note that without using any prior on $H_0$, we have $H_0 = 71.9\pm 4 $ km s${}^{-1}$ Mpc${}^{-1}$ and 
$H_0 = 74.7\pm 5$ km s${}^{-1}$ Mpc${}^{-1}$ both at 68\% C.L. from CMB + BAO and CMB + BAO + LSS, respectively. 
With the introduction of HST in the analysis, we have $H_0 = 72.8\pm 1.7 $ km s${}^{-1}$ Mpc${}^{-1}$ 
and $H_0 = 73.3\pm 1.7$ km s${}^{-1}$Mpc${}^{-1}$  both at 68\% C.L. from CMB + BAO + HST and joint analysis using full data.
Therefore, with all the data sets (with or without using the local prior on $H_0$) the DM-photon coupling model favors the local measurement 
$H_0=73.24\pm 1.74$ km s${}^{-1}$ Mpc${}^{-1}$. 
Thus, the DM-photon coupling model developed here can serve as an alternative to explain the  well-known tension of $H_0$.
Other perspectives on alleviating the tension on $H_0$ are investigated in \cite{H01,H02,H03,H04,H05,H06a,H06b,H07,H08,H09,H10}.

Another cosmological tension arises from the predictions of the direct measurements of LSS and CMB for $\sigma_8$.
The results from Planck CMB yield the value of amplitude of matter density fluctuations, $\sigma_8= 0.831\pm 0.013$ \cite{Planck2015},
which is about $2\sigma$ higher than $\sigma_8= 0.75\pm 0.03$ as given by the Sunyaev-Zel’dovich cluster abundance measurements \cite{p14}. 
From Table \ref{Table_M1}, we note that the  constraints on $\sigma_8$ from CMB + BAO + LSS and joint analysis using the full data are 
$\sigma_8= 0.781\pm 0.028$ and $\sigma_8= 0.778\pm 0.027$ at 95\% C.L., respectively. These, in addition to the common region of the contours in the LSS 
cases with the vertical light red band in Fig. \ref{PS_model1}, indicate some consistency of the DM-photon coupling model based $\sigma_8$ values 
with the local measurements  like Sunyaev-Zel’dovich cluster abundance measurements, galaxy cluster count and weak gravitational lensing etc. 
Thus, the DM-photon coupling model alleviates the $\sigma_8$ tension as well to some extent when LSS data is included in the analysis.

The upper bound on the mass scale of active neutrinos is obtained as $\sum m_{\nu} < 0.38$ eV with minimal data set CMB + BAO, whereas a higher 
upper bound $\sum m_{\nu} < 0.83$ eV is obtained with the full data set, both at 95\% C.L. We see that the constraint on $\sum m_{\nu}$ 
becomes weaker by a factor of two in the case of full data set. The presence of massive neutrinos reduces the growth of perturbations 
(reducing the growth rate of structures) below their free-streaming length. 
Thus, the data from LSS (prior in the $\sigma_8 - \Omega_m$ plane) which are physically dependent on small scale approximation, 
and where massive neutrinos play an important role, are responsible for this degeneracy (double the neutrino mass scale in our analysis). 
It is difficult to quantify the individual physical effects responsible for the constraints on $\sum m_{\nu}$, 
knowing that many systematic effects (calibration of the mass observable relation, modeling the number of halos, etc.) 
can affect LSS data. However, evidently the constraints $\sum m_{\nu} < 0.97$ eV and $\sum m_{\nu} < 0.83$ eV when LSS data are added in 
the analysis are related to the tension on $\sigma_8$. These constraints differ up to 
2$\sigma$ C.L. from the ones obtained using CMB + BAO and CMB + BAO + HST data. This difference about 2$\sigma$ CL is also transmitted to the 
neutrino mass scale.

Further, we notice that the spectrum index ($n_s$) of the primordial scalar perturbations is compatible with 
a scale invariant spectrum ($n_s =1$) at 2$\sigma$ C.L. in all the four cases of our analysis here. In the case CMB + BAO + LSS + HST,
$n_s =1$ is compatible at 1.5$\sigma$ C.L. Taking the minimal $\Lambda$CDM model, Planck team \cite{Planck2015_inflation} 
has ruled out $n_s =1$ at 5.6$\sigma$ C.L. Therefore, the extended model presented here (DM-photon coupling + $N_{\rm eff}$) 
can reduce it up to 4$\sigma$ C.L. Effects of some extended scenarios on $n_s$ are also discussed in \cite{Planck2015_inflation}.
A scale invariant spectrum is investigated in the light of $H_0$ and $\sigma_8$ tensions in \cite{H04}.

The decay process of DM into photons represents a direct nonconservation in total radiation energy density throughout the cosmic history. Thus, it is also
expected that this process can also change $N_{\rm eff}$ and $r_{\rm drag}$ (BAO acoustic scale at drag epoch).  
These quantities directly depend on $\rho_{\gamma}$. Therefore, the effective number of species can be parametrized (when the neutrinos are relativistic) by
\begin{eqnarray}
 N_{\rm eff} =  \frac{8}{7} \left(\frac{4}{11} \right)^{-\frac{4}{3}} \left(\frac{\rho_{\nu}}{\rho_{\gamma}} \right).
\end{eqnarray}

As we have a change over $\rho_{\gamma}$ from the standard evolution prediction, the nonconservation on photon density can influence
$N_{\rm eff}$ to nonstandard values. Also, a DM-photon coupling can affect the BAO acoustic scale at drag epoch $r_{\rm drag}$, 
since this quantity depends tightly on the baryon-photon ratio $R = 3 \rho_b/ 4 \rho_{\gamma}$,
\begin{eqnarray}
\label{r_drag}
 r_{\rm drag} (z_{\rm drag}) = \int_{z_{\rm drag}}^{\infty} \frac{c_s(z)}{H(z)} dz, \, \, {\rm with} \,\, c_s(z)= \frac{c}{\sqrt{3 + 3R(z)}},
\end{eqnarray}
where $\rho_{\rm b}$ stands for the baryon density and $c_s$ is the sound velocity as a function of the redshift. Besides the nonconservation 
of the photons induced by $\Gamma_{\gamma }$, the addition of $w_{\rm de}$ and $N_{\rm eff}$ is expected to cause variations on 
$r_{\rm drag}$ too.

\begin{figure*}\centering
\includegraphics[width=7.0cm]{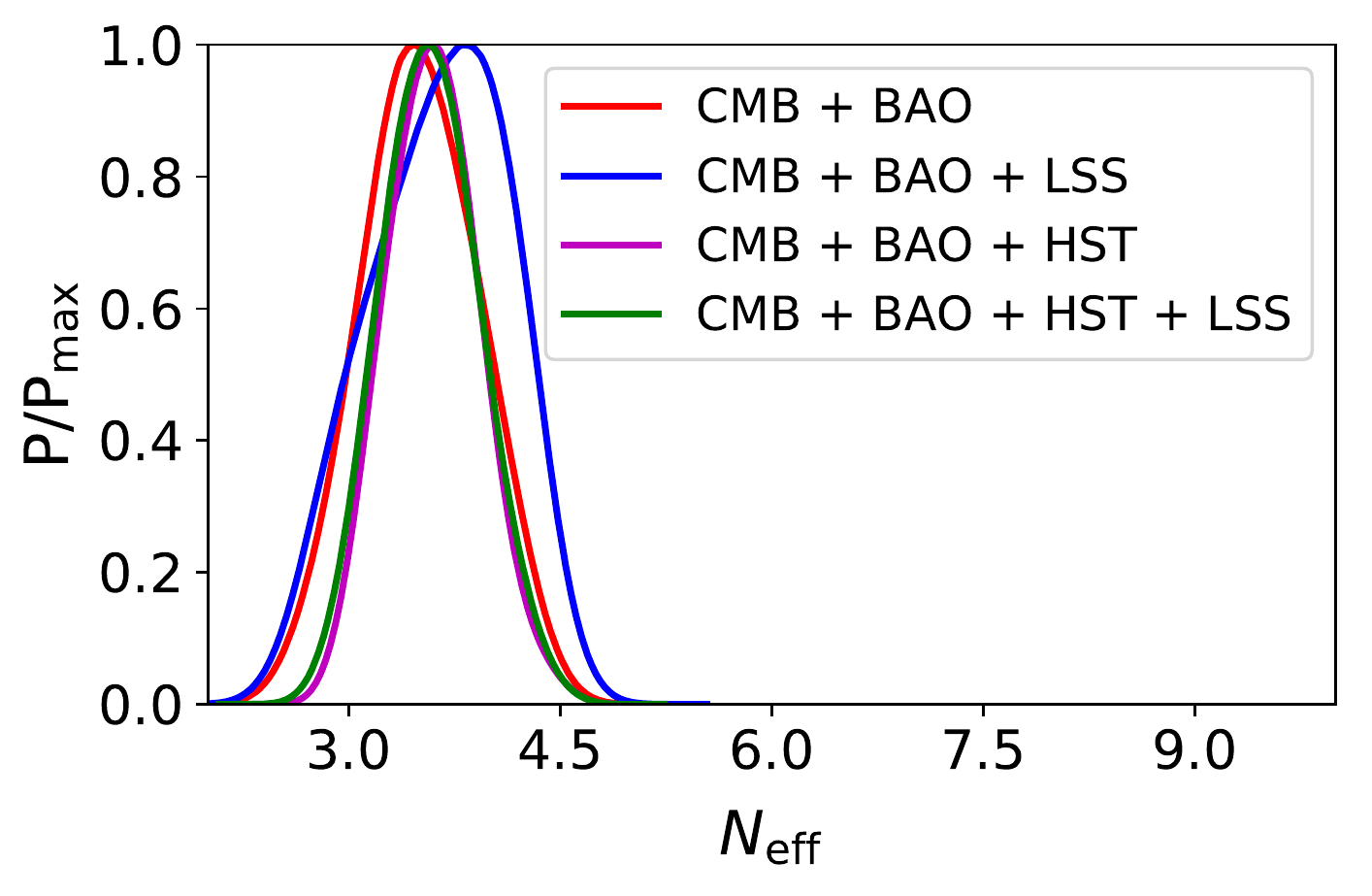}
\includegraphics[width=7.0cm]{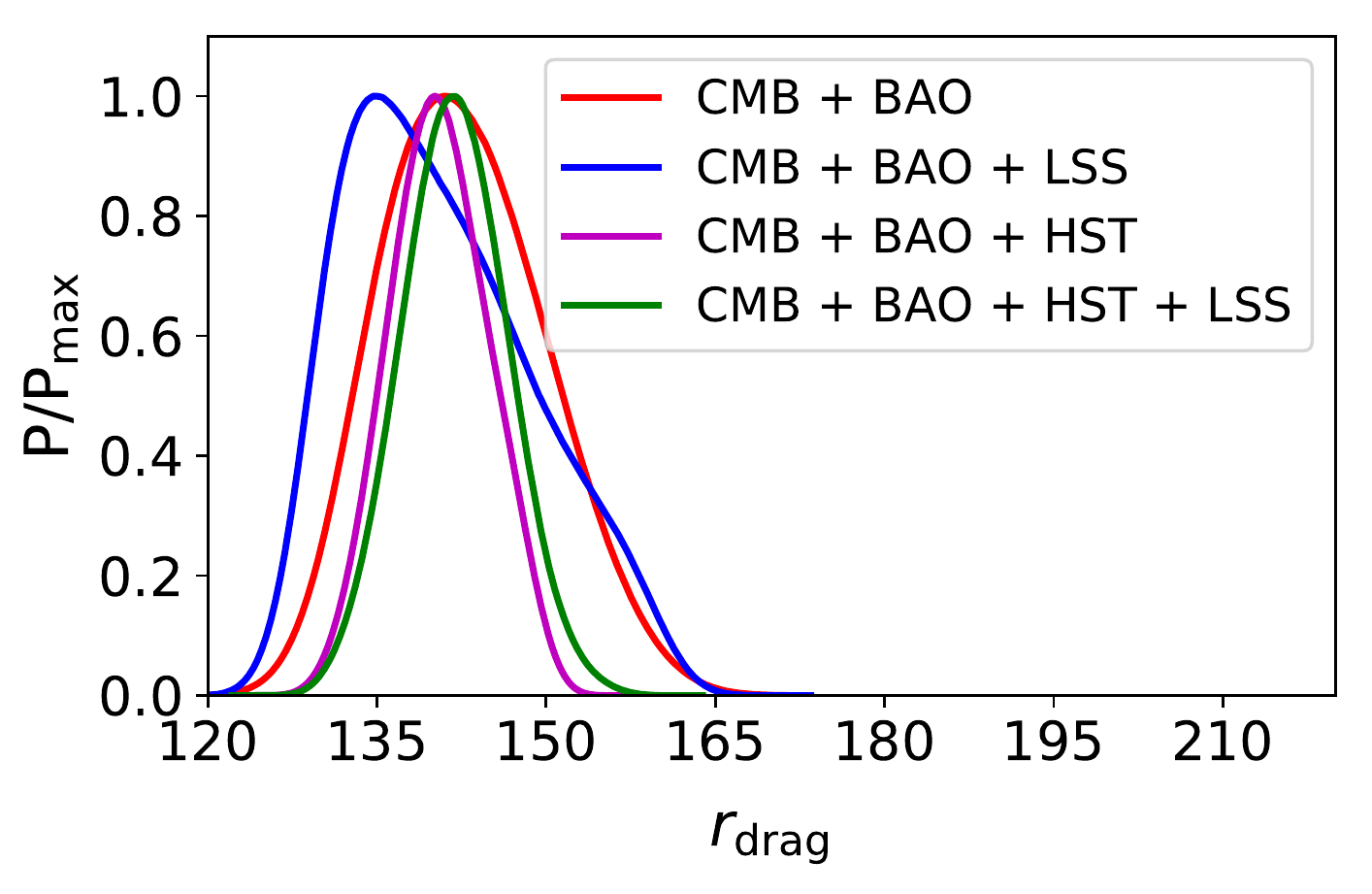}
\caption{\label{NR_model2} {One-dimensional marginalized distribution of $N_{\rm eff}$ and $r_{\rm drag}$ (measured in Mpc).}}
\end{figure*}

The Planck team \cite{Planck2015} has reported $r_{\rm drag} = 147.60 \pm 0.43$ Mpc from TT + lowP + lensing in the minimal base $\Lambda$CDM model. 
The authors in \cite{rs01} within $\Lambda$CDM + $N_{\rm eff}$ have obtained $r_{\rm drag} = 143.53 \pm 3.3$ Mpc. Using only 
low-redshift standard ruler, the constraint $r_{\rm drag} = 143.9 \pm 1.9$ Mpc is obtained in \cite{rs02}. 
As already mentioned, the three free parameters ($\Gamma_{\gamma}$, $w_{\rm de}$ and $N_{\rm eff}$) in our model can affect 
$r_{\rm drag}$. Figure \ref{NR_model2} shows the one-dimensional marginalized distribution on $N_{\rm eff}$ and $r_{\rm drag}$ (measured in Mpc)
from the four data combinations used in this work. We find $r_{\rm {drag} } = 142.5^{+6.8}_{-8.2}$ Mpc at 68\% C.L. from CMB + BAO. 
This value is compatible with all the measures mentioned above at 68\% C.L. The other values, from the other combinations, 
are also compatible to the values presented in \cite{Planck2015, rs01, rs02} at 68\% C.L. However, due 
to the our extended model and the used data combinations, the most reasonable comparison on $r_{\rm {drag}}$ is made for CMB + BAO only. 
Within our model and analysis, we find that the best fit for $N_{\rm eff}$ can deviate significantly from its standard value 
($N_{\rm eff} = 3.046$) in all the cases analyzed here, but when the borders are observed even at 68\% C.L., no evidence for dark radiation is
found. In general, the DM-photon coupling can significantly change the best fit on $N_{\rm eff}$ and $r_{\rm drag}$, 
due to nonconservation of photons through the expansion of the Universe, but the observational bounds 
are compatible at 68\% C.L. with the standard results of the $\Lambda$CDM model.



\section{Conclusion}

In this work, we have investigated an extension of the $\Lambda$CDM model where a nonminimal DM-photon coupling is assumed, and the dark energy 
is characterized by a constant equation of state parameter. In addition, massive neutrinos are also considered. 
We have observed significant effects of the considered cosmological scenario on the CMB TT and matter power spectra (see Sec. \ref{powerspectra}).
We have obtained observational constraints on the DM-photon coupling model parameters using the latest data from CMB, BAO, HST and LSS. In particular,
we have found the upper bound on the coupling parameter, viz.,  $\Gamma_{\gamma } \leq 1.3 \times10^{-5}$
using the full data set CMB + BAO + HST + LSS. As a direct consequence of the decay of DM into photons, we have found that this scenario 
can naturally solve the current tension on the Hubble constant by providing high values of $H_0$ consistent with the local measurements. 
Thus, alteration in the standard dynamics of photons through cosmic expansion can be a viable alternative to describe physics beyond the 
$\Lambda$CDM model in light of the current observational tensions. Here the cosmological model with DM-photon coupling has proved to be able to 
resolve the tension on $H_0$. A step forward within this perspective could be to consider a curvature-photon coupling  and investigate 
its cosmological consequences. This can serve as a test to probe the modified gravity on the CMB scale, and reconcile the current tension between
LSS and CMB as well.

\section*{Acknowledgments}
S.K. gratefully acknowledges the support from SERB-DST project No. EMR/2016/000258. S.K.Y. acknowledges the Council of Scientific \& Industrial
Research (CSIR), Govt. of India, for awarding Junior Research Fellowship (File No. 09/719(0073)/2016-EMR-I). 


\end{document}